# High-Throughput ab-initio Dilute Solute Diffusion Database

Henry Wu[1], Tam Mayeshiba[2], Dane Morgan[1]

1. Department of Materials Science and Engineering, University of Wisconsin-Madison, Madison, USA
2. Materials Science Program, University of Wisconsin-Madison, Madison, USA

Corresponding author: Dane Morgan (ddmorgan.wisc.edu)

**Abstract:**

We demonstrate automated generation of diffusion databases from high-throughput density functional theory (DFT) calculations. A total of more than 230 dilute solute diffusion systems in Mg, Al, Cu, Ni, Pd, and Pt host lattices have been determined using multi-frequency diffusion models. We apply a correction method for solute diffusion in alloys using experimental and simulated values of host self-diffusivity. We find good agreement with experimental solute diffusion data, obtaining a weighted activation barrier RMS error of 0.176 eV when excluding magnetic solutes in non-magnetic alloys. The compiled database is the largest collection of consistently calculated *ab-initio* solute diffusion data in the world.

**Background & Summary:**

Solute diffusion is the way in which impurities are transported in alloys, and many important material properties depend critically upon this transport, such as phase transition kinetics[1-3]. In general solute diffusion is controlled by the random jumps of point defects within the material. In the case of vacancy mediated diffusion in dilute solid solution alloys, the impurity diffusion coefficient can be accurately predicted from the rates of atomic vacancy exchanges around the impurity, and robust formulae have been developed for major crystal structures[4].

Despite the importance of impurity diffusion coefficients, only a small fraction of dilute binary alloy diffusivities have been experimentally measured[5,6]. The limited data is due to many experimental challenges, including a lack of corresponding radioactive tracer, detection limitations for slow diffusers, and metastability of the host crystal structure, as well as simply the time and cost of exploring the tens of thousands of possible systems. First-principles theoretical methods overcome these issues, as they are able to utilize a wide variety of elemental species, sample and quantify high activation barriers, work with metastable crystal structures, and can be performed relatively cheaply and quickly compared to experiments when properly automated. A computational approach is also able to provide the diffusion data in a consistent framework, allowing all diffusivities to be compared on equal footing.

Expanding upon previous theoretical studies of dilute solute diffusion in alloys[7-14], we present in this work the largest consistently calculated *ab-initio* solute diffusion database to-date. This database consists of more than 230 dilute solute diffusion systems in Mg, Al, Cu, Ni, Pd, and Pt hosts. These diffusion calculations were automated using our high-throughput workflow software, the MAterials Simulation Toolkit (MAST),[15-17] developed at the University of Wisconsin-Madison. MAST is built upon pymatgen[18] and automatically handles input/output processing of *ab-initio* calculations and manages job submission to cluster queues. MAST can be used to control complex workflows, and was used here to manage multifrequency model calculations on a large number of systems.

The paper is organized as follows. We first briefly outline our computational methodology for generating dilute solute diffusion data and detail our empirical corrections. An overview of the structure and description of the data will then be presented. Finally we demonstrate the validity of our data with an analysis of associated DFT errors and comparisons to experimental diffusion measurements.

## Methods:

### Computational methods

We perform all calculations using the Vienna *ab-initio* Simulation Package (VASP)[19-22]. We treat exchange–correlation in the Generalized Gradient Approximation (GGA), as parameterized by Perdew, Burke, and Ernzerhof (PBE)[23,24]. The projector augmented wave method (PAW)[25,26] pseudopotentials were used with a plane wave cutoff of 350 eV for all systems. The constant 350 eV energy cutoff was used to keep consistency and is higher than the largest ENMAX of elements calculated. Bulk and defect calculations were done using 4×4×3 HCP conventional supercells for Mg alloys containing 96 atoms and 3×3×3 cubic FCC supercells for Al, Cu, Ni, Pd, and Pt alloys containing 108 atoms. The Brillouin zone was sampled by a 5×5×5 Gamma centered mesh for the HCP supercells and a 4×4×4 Monkhorst-Pack k-point mesh for the FCC supercells. Errors in energy are converged to less than 1 meV/atom with respect to the energy cutoff and k-points; errors in force are relaxed to less than 0.01 eV/Å. All runs that require magnetization were done as spin-polarized calculations; these include all Ni alloys, and Cr, Mn, Fe, Co, and Ni solutes. The need to run spin-polarized calculations for magnetic solutes in non-magnetic hosts has previously[8,11] been found to be essential for diffusion calculations. Additional computational method effects such as finite supercell errors and comparison between different exchange-correlation functionals will be discussed in the validation section.

Migration barriers for atomic jumps were calculated using the climbing image nudged elastic band (CI-NEB) method with a single intermediate image. For the transitions we consider, which are single atom jumps to nearest neighbor sites, a single image is sufficient to determine the transition saddle point. Migration attempt frequencies ($\upsilon_{hop}$) were calculated with the Vineyard[27] approach. However, rather than computing all $3n$ vibrational modes, we consider only the vibrational modes of the hopping atom (with all other atoms held fixed) in its initial position ($\upsilon^{initial}$) and at the saddle point configuration ($\upsilon^{saddle}$):

$$\nu_{hop} = \frac{\prod_1^{3n} \nu_i^{initial}}{\prod_1^{3n-1} \nu_i^{saddle}} \sim \frac{\prod_1^{3} \nu_i^{initial}}{\prod_1^{2} \nu_i^{saddle}}.$$

### Dilute solute diffusion models

We calculate solute diffusion coefficients by following the multi-frequency framework developed by LeClaire[28], using the five-frequency diffusion model[1,4] for FCC (Figure 1a) and the eight-frequency diffusion model[29] for HCP (Figure 1b). These diffusion models assume dilute solute concentrations and therefore do not include solute-solute interactions. Each jump frequency ($\omega_i$), is calculated from DFT migration barriers ($E_i$) and attempt frequencies ($\upsilon_i$) in the simple Arrhenius expression

$$\omega_i = \nu_i \exp\left(\frac{-E_i}{k_B T}\right),$$

where $k_B$ is the Boltzmann constant and $T$ is the temperature. In the five-frequency FCC diffusion model, $\omega_0$ is the bulk vacancy hop rate away from any solutes, $\omega_1$ is the vacancy-solute rotation hop, $\omega_2$ is the vacancy-solute exchange hop, and $\omega_3$ and $\omega_4$ are the vacancy-solute dissociation and association hops, respectively. In the eight-frequency HCP diffusion model, $\omega_a$ and $\omega'_a$ are the vacancy-solute rotation hops from basal orientation to c-axis and vice versa, $\omega_b$ and $\omega'_b$ are the vacancy-solute rotation hops within the basal and c-axis planes, $\omega_c$ and $\omega'_c$ are the vacancy-solute dissociation hops from the basal and c-axis configurations, and $\omega_X$ and $\omega'_X$ are the vacancy-solute exchange hops within the basal and c-axis planes. For the FCC systems, the prefactors for all five frequencies were calculated and included. For the HCP systems, two prefactors were calculated and used, one for all solute atom transitions ($\omega_X$ and $\omega'_X$) and one for all solvent atom transitions ($\omega_a$, $\omega'_a$, $\omega_b$, $\omega'_b$, $\omega_c$, and $\omega'_c$).

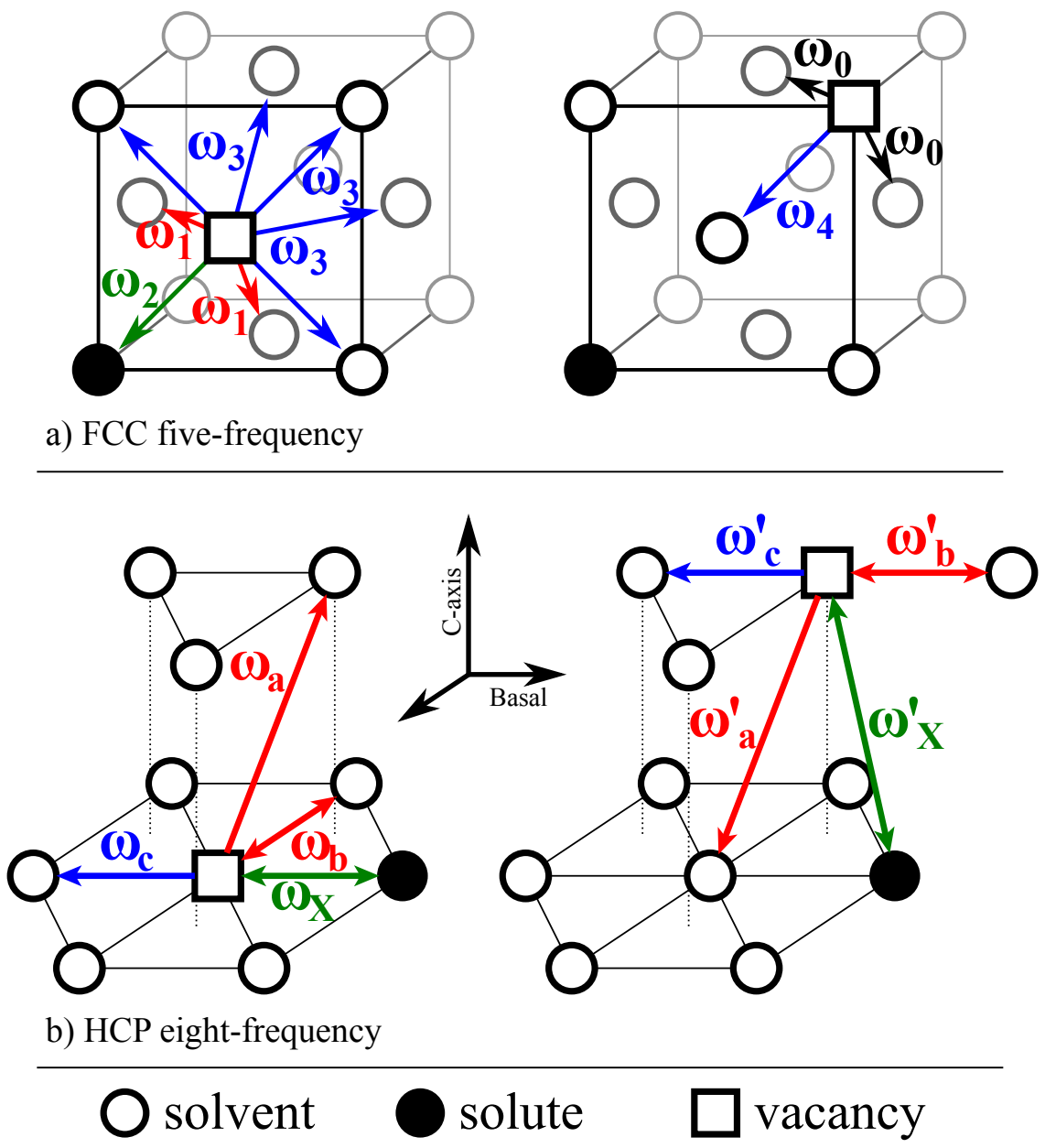

Figure 1: a) Atomic jumps required for the FCC five-frequency diffusion model, b) atomic jumps required for the HCP eight-frequency diffusion model.

To improve the predictive capabilities of DFT diffusion, we propose a correction on top of direct DFT calculated solute diffusivity, by scaling according to how much the DFT host self-diffusivity deviates from the experimental self-diffusivity. We accomplish this by multiplying the raw DFT diffusivities by a correcting Arrhenius equation,

$$D_{corrected}^{solute} = A_{shift} \exp\left(\frac{-E_{shift}}{k_B T}\right) \cdot D_{DFT}^{solute},$$

where the correctional shift parameters, $A_{shift}$ and $E_{shift}$, are determined by fitting the DFT host self-diffusivity to experimental measured self diffusivity such that,

$$D_{experiment}^{host} \sim A_{shift} \exp\left(\frac{-E_{shift}}{k_B T}\right) \cdot D_{DFT}^{host}.$$

Table I reports these correction parameters for all six host elements along with the final corrected diffusion constant and activation barrier. All solute diffusivities and diffusion parameters reported will be values after this corrective procedure. This correction is not essential but improves results compared to experiments and creates almost no loss of generality for our approach because self-diffusion coefficients are known for almost all the elements of interest.

Table I: Correctional shifts, $A_{shift}$ and $E_{shift}$, for DFT predicted Ag, Cu, Ni, Pd, Pt, and Mg self-diffusivity fitted from experimental diffusion[6]. The corrected DFT self-diffusion constant, $D_0$, and activation barrier, $Q$, are also reported.

| | $A_{shift}$ | $E_{shift}$ [eV] | $D_0$ [cm$^2$/s] | Q [eV] |
|---|---|---|---|---|
| **Al** | 12 | 0.20 | 0.065 | 1.266 |
| **Cu** | 80 | 0.40 | 0.282 | 2.080 |
| **Ni** | 500 | 0.47 | 2.145 | 2.954 |
| **Pd** | 20 | 0.55 | 0.072 | 2.646 |
| **Pt** | 20 | 0.85 | 0.062 | 2.676 |
| **Mg** | 240 | 0.20 | 1.362 | 1.406 |

**Code availability**

The MAterials Simulation Toolkit (MAST)[15,17] is the code package used for the calculation of these diffusion coefficients. MAST is an open-source code released with the Massachusetts Institute of Technology (MIT) license and is freely accessible at https://pypi.python.org/pypi/MAST.

## Data Records

The full diffusion dataset is publically available at Figshare (see Data Citation 1: Figshare http://dx.doi.org/10.6084/m9.figshare.1546772) and at our own interactive web page[30] (http://diffusiondata.materialshub.org). The data for each host element catalogs the various properties of the host element, hopping properties of the solute in the host, and extracted solute diffusion parameters. There is only one set of host element properties, while additional data columns are used for each additional solute element. The solute diffusion parameters, solute diffusion constant, $D_0$ and solute diffusion activation energy, $Q$, can be used in the following Arrhenius diffusion equation to generate the temperature, $T$, dependent solute diffusivity:

$$D^{solute}(T) = D_0^{solute} \exp\left(\frac{-Q^{solute}}{k_B T}\right). \quad (1)$$

**Graphical representation of the results**

In Figure 2 we plot the DFT diffusion activation energies in each of the six host alloys. These diffusion activation barriers are extracted from our DFT diffusivities in the temperature range between the host element's melting temperature and half melting temperature. Quantitative similarities can be seen between the 3d, 4d, and 5d solutes, with a noticeable dip for the 3d magnetic elements, Cr, Mn, Fe, Co, and Ni. This dip is because while the host Mg, Al, or Cu does not show any magnetization; the presence of some of these magnetic solutes induces a moment at the transition state of the solute-vacancy exchange. This effect reduces the energy barrier for those transitions, resulting in the dips seen in Figure 2. If these solutes were calculated without spin-polarization, the 3d curves would instead follow the same trend as the 4d and 5d curves.

An increase in the diffusion activation energy correlates with an increased d-shell filling, peaking near half d-filling, and then finally decreasing back down as the d-shell completely fills. This smooth change is only broken by the above-mentioned magnetic 3d solutes. The amount of change in the activation energy becomes more significant at higher d-shells, with larger barrier changes in 5d as compared with 3d when moving across the table. Between different d-shells, diffusivities converge and cross over near the Ti/V groups on the left and near the Ni/Cu groups on the right. These transition points are not surprising as elements in these periodic groups are quite similar chemically. The resulting effect gives higher activation energies with higher d-shell within the range between the Ti/V and Ni/Cu groups, and lower activation energy with higher d-shells outside of this range.

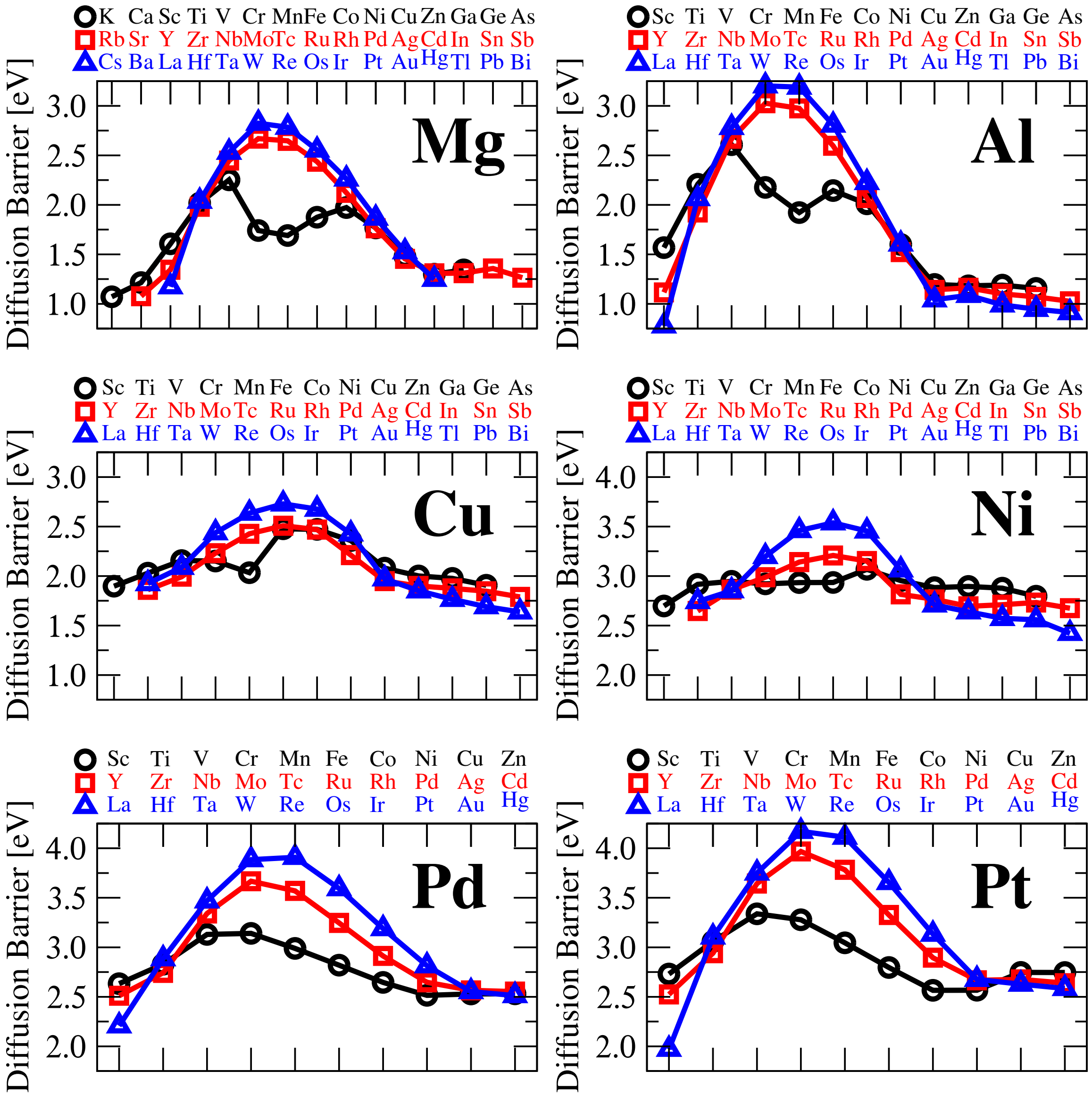

Figure 2: Trend in solute diffusion activation barriers in all host alloys, Mg, Al, Cu, Ni, Pd, and Pt from DFT calculations across the periodic table. The barriers are extracted from the temperature range between the host element's melting temperature and half melting temperature. For Mg, only the basal diffusion barrier is plotted; the trend for the c-axis diffusion barrier is almost the same.

## Technical Validation

### Validation with experimental diffusion measurements

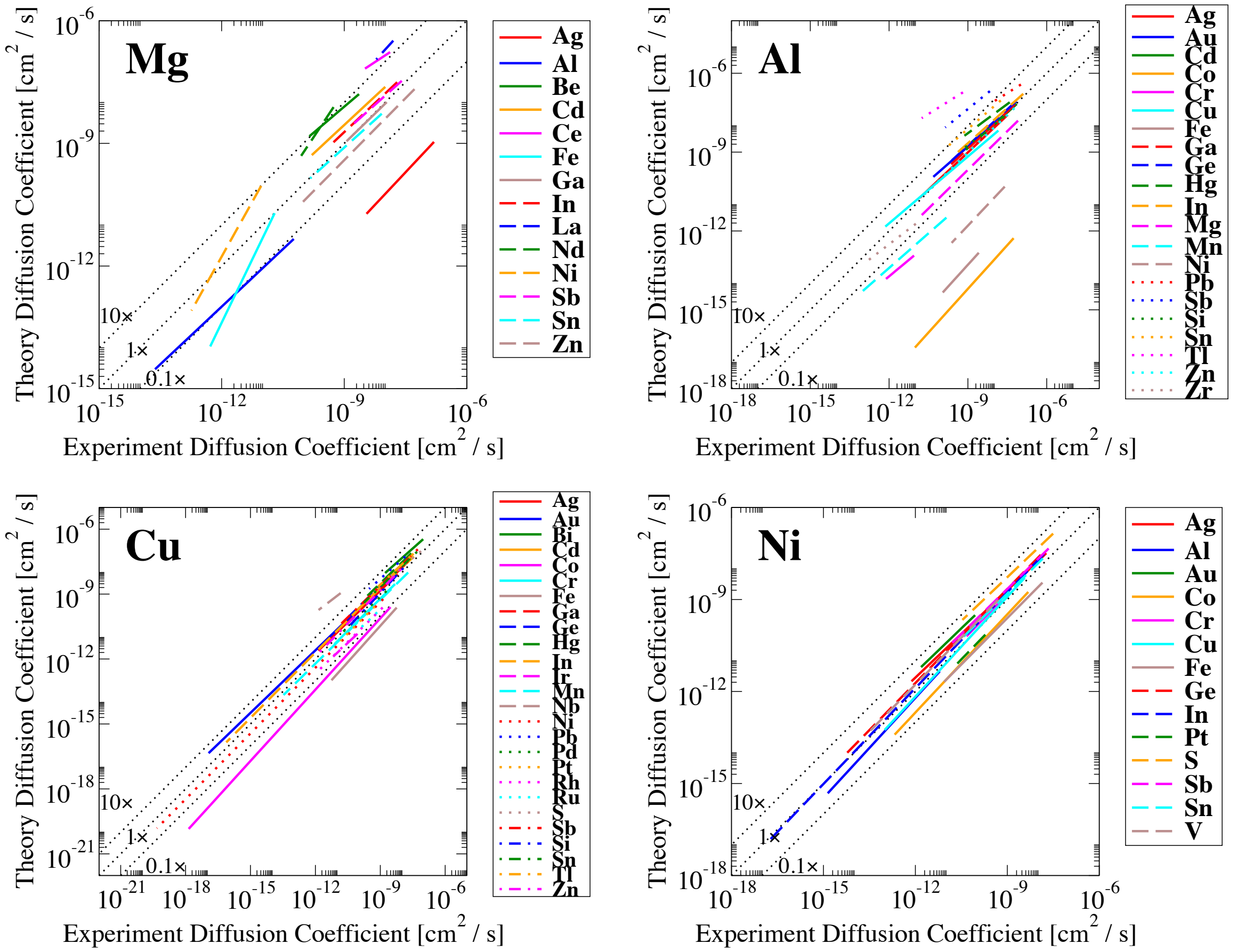

Figure 3: Comparison between DFT solute diffusivities and experimental measurements. Each line represents a solute in Mg, Al, Cu, or Ni. The DFT diffusivity for each solute is matched up with the experimental diffusivity within the experimental temperature range. Dotted black lines represent the identity line (1×) along with 10× and 0.1× DFT vs. experiment.

Figure 3 compares corrected DFT diffusion values to experimentally measured diffusion coefficients for dozens of dilute solutes in Mg, Al, Cu, and Ni. In these plots, the DFT diffusivity is shown for the same temperature range as used in the experimental data. Both experimental and DFT values are determined from Arrhenius fits (Eq. (1)) to the exact measurements and calculations. The experimental and DFT values for a given system and temperature are then viewed as an (x,y) pair and plotted. We connect these points with lines since Arrhenius expression trends are perfectly linear on log-log plots. Perfect agreement would result in a 45° y=x line, right along the diagonal. A line that is shifted by a constant off the central diagonal represents a multiplicative factor between theory and experiment, i.e., a discrepancy in $D_0^{solute}$ in Eq. (1). Lines that are not on a 45° slope indicate activation barrier differences between theory and experiment, i.e., a discrepancy in $Q^{solute}$ in Eq. (1). More than half of all solutes in Al and almost all solutes in Mg, Cu, and Ni fall within a factor of 10 with respect to the experiment. The largest diffusivity disagreement is seen for solute diffusion in Al, where DFT over-predicts Tl diffusion by three orders of magnitude and under-predicts Co and Fe diffusion by four orders of magnitude each. In Mg, the solute Ag is under-predicted by DFT, while the largest barrier disagreement is found for Fe and Ni. It is clear that most of the solutes that show large disagreement between theory and experiment are the magnetic elements, Cr, Mn, Fe, Co, and Ni. The close agreement we find for all solutes in Ni, which were all done spin-polarized, suggests that this is not an intrinsic failure for all magnetic calculations. We instead conclude that the issue lies with the configuration of a single solute magnetic moment surrounded by host atoms with no moments. Either DFT is not able to capture all the effects of this interaction, or some other diffusive mechanism is activated by this single atom moment.

We quantify the DFT/experimental agreement using three host-dependent metrics: two solute diffusion barrier RMS errors, for both weighted and unweighted averages, and a solute diffusion coefficient ratio.

The unweighted diffusion barrier RMS error is calculated as:

$$E_{host}^{RMS}\,[eV] = \sqrt{\frac{\sum_{i=1}^{n}\left[\left(E_i^{DFT} - E_i^{expt}\right)^2\right]}{n}},$$

while the weighted diffusion barrier RMS error is computed as:

$$E_{host}^{w-RMS}\,[eV] = \sqrt{\frac{\sum_{i=1}^{n}\left[\left(\frac{1}{T_i^{low}} - \frac{1}{T_i^{high}}\right)\cdot\left(E_i^{DFT} - E_i^{expt}\right)^2\right]}{\sum_{i=1}^{n}\left(\frac{1}{T_i^{low}} - \frac{1}{T_i^{high}}\right)}},$$

where $E_i^{DFT}$ and $E_i^{expt}$ are the DFT and experimental diffusion barriers for solute *i*, respectively, while $T_i^{low}$ and $T_i^{high}$ form the experimental temperature range in Kelvin for solute *i*, and *n* is the number of solutes compared. This method places lower weights for narrower experimental temperature ranges due to the intrinsically higher fitting error on the experimental diffusion. $E_{host}^{RMS}$ and $E_{host}^{w-RMS}$ represent the diffusion activation barrier RMS error in units of eV for a particular host system, unweighted and weighted, respectively.

The diffusion coefficient ratio metric is the average of the log of ratios of DFT to experimental *D* values, which is computed in the following manner:

$$\log_{10}\left[D_{host}^{ratio}\right] = \frac{\sum_{i=1}^{n}\left|\log_{10}\left[D_i^{DFT}/D_i^{expt}\right]\right|}{n},$$

where $D_i^{DFT}$ and $D_i^{expt}$ are average DFT and experimental diffusion coefficients for solute *i*, over the experimental measurement range. $D_{host}^{ratio}$ represents an average deviation factor between DFT and experiment for a particular host system. Please note that the number given is not for the log deviation error, rather it is a direct diffusion ratio factor $D_{host}^{ratio}$. From Figure 3 we find this metric triplet, ($E_{host}^{RMS}$, $E_{host}^{w-RMS}$, $D_{host}^{ratio}$), to be: (0.404eV, 0.436eV, 5.44) for Mg-host, (0.294eV, 0.229eV, 14.7) for Al-host, (0.183eV, 0.134, 3.32) for Cu-host, and (0.130eV, 0.134eV 2.30) for Ni-host. Combining all experimental comparisons for these four hosts, we find our performance metric, ($E_{host}^{RMS}$, $E_{host}^{w-RMS}$, $D_{host}^{ratio}$), to be: (0.264eV, 0.231eV, 5.16). Excluding the magnetic solutes from non-magnetic hosts, our performance metric improves to: (0.225eV, 0.176eV, 3.31).

**Analysis of associated computational errors**

To quantify the limitations of our computational methodology, we compute the errors resulting from several aspects of our calculation settings. These include finite-size supercell effects, choice of the exchange-correlation functional, effect of extended solute-vacancy binding, and approximation of the hopping atom attempt frequency.

DFT calculations are widely used because of their efficiency, reliability and transferability. However, they are still generally limited to calculations of less than about 1000 atoms, and typically many fewer for studies involving thousands of calculations. The small periodic supercell sizes can introduce significant finite size cell effects due to strain and other fictitious image effects, and must be carefully considered. We estimate the magnitude of this effect by calculating the vacancy formation and migration energy for Mg with 3×3×2 (36 atoms), 4×4×3 (96 atoms), and 6×6×4 (288 atoms) HCP supercells, and for Pd/Pt with 2×2×2 (32 atoms), 3×3×3 (108 atoms), and 4×4×4 (256 atoms) FCC supercells. We then fit a linear relation between these energies versus the inverse of the total number of atoms at each size. We find that Mg vacancy formation energy is almost independent with respect to system size, while both Pd and Pt vacancy formation energies decrease with system size. The extrapolated formation energy at infinite size, corresponding to the y-intercept of the fit, is within 50 meV of that from the size we use for all future diffusion calculations (4×4×3 for HCP, and 3×3×3 for FCC). The extrapolated vacancy migration energy at infinite size is within 30 meV to that from the size we use. For the smallest Mg supercell size, 3×3×2, we find that only two unit cells in the c-axis direction is clearly insufficient, as the c-axis vacancy migration energy deviates significantly from linear scaling.

In Kohn-Sham DFT, the exchange-correlation (xc) functional is an approximation to the exact exchange interaction and electronic correlation between many-body electrons. Approximating the xc functional is necessary because the exact functional form is unknown. No current xc functional is accurate for all system properties, and a variety of functionals should be tested for the application of interest. We test the vacancy formation and migration energies of the six host elements against experimental measurements for four different xc functionals: local density approximation (LDA), Perdue-Wang'91 (PW91), Perdew-Burke-Ernzerhof (PBE), and PBE solid (PBEsol). All of these are widely used exchange-correlation functionals in DFT.

Table II: Predicted vacancy formation $V_{form}$ (eV) and vacancy migration $V_{mig}$ (eV) energies for the six host elements, Al, Cu, Ni, Pd, Pt, and Mg. Different DFT exchange-correlation functionals, PBE, LDA, PW91, and PBEsol, are compared against experimental measurements. The migration energy for Mg is an average value of the basal and c-axis diffusivities.

| | **Al** | | **Cu** | | **Ni** | | **Pd** | | **Pt** | | **Mg** | |
|---|---|---|---|---|---|---|---|---|---|---|---|---|
| | $\mathbf{V_{form}}$ | $\mathbf{V_{mig}}$ | $\mathbf{V_{form}}$ | $\mathbf{V_{mig}}$ | $\mathbf{V_{form}}$ | $\mathbf{V_{mig}}$ | $\mathbf{V_{form}}$ | $\mathbf{V_{mig}}$ | $\mathbf{V_{form}}$ | $\mathbf{V_{mig}}$ | $\mathbf{V_{form}}$ | $\mathbf{V_{mig}}$ |
| **PBE** | 0.485 | 0.581 | 0.963 | 0.717 | 1.645 | 0.957 | 1.137 | 0.959 | 0.611 | 1.215 | 0.798 | 0.408 |
| **LDA** | 0.580 | 0.603 | 1.269 | 0.830 | 1.587 | 1.078 | 1.407 | 1.127 | 0.878 | 1.425 | 0.802 | 0.415 |
| **PW91** | 0.461 | 0.538 | 1.025 | 0.698 | 1.333 | 0.930 | 1.113 | 0.916 | 0.608 | 1.169 | 1.196 | 0.396 |
| **PBEsol** | 0.632 | 0.606 | 1.249 | 0.805 | 1.580 | 1.059 | 1.363 | 1.084 | 0.840 | 1.393 | 0.831 | 0.413 |
| **Expt.**[6] | 0.67±0.03 | 0.61±0.03 | 1.28±0.05 | 0.70±0.02 | 1.79±0.05 | 1.04±0.04 | 1.70, 1.85 | 1.03±0.3 | 1.35±0.05 | 1.43±0.05 | 0.58-0.81 | 0.45-0.6 |

In Table II, we show the predictions of the vacancy formation and migration energies from PBE, LDA, PW91, and PBEsol for Al, Cu, Ni, Pd, Pt, and Mg. From the data, there is no clear functional which perform significantly better than others. For the elements calculated, almost all xc functionals come close to matching the experimental vacancy migration energy, while more deviations are seen for vacancy formation, especially for Pd and Pt. Since the activation barrier for self-diffusion by a vacancy mechanism is simply the sum of vacancy formation and migration, these results suggest that all tested xc functionals still deviate by several hundreds of meV compared to experimental diffusion barriers. Because we apply the self-diffusivity correction onto all solute diffusivity results, there is little difference between each of these xc functionals, and we chose to use the PBE xc functional for all our solute diffusion calculations.

Within the five-frequency model, $\omega_3$ and $\omega_4$ represent the dissociation and association hops between a solute and vacancy, respectively. This diffusion model assumes only first nearest-neighbor (1NN) interactions between the solute and vacancy, meaning that all energy changes for vacancy movement away from the 1NN configuration are equivalent, whether it be to the second (2NN), third (3NN), or fourth (4NN) nearest-neighbor. The assumed complete dissociation beyond 1NN also allows the difference in energy barrier between $\omega_3$ and $\omega_4$ to act as the solute-vacancy binding energy within the diffusion model. However, since the solute-vacancy interactions in real systems do not stop at 1NN, the magnitude of further neighbor binding and their effect on solute diffusion must be considered.

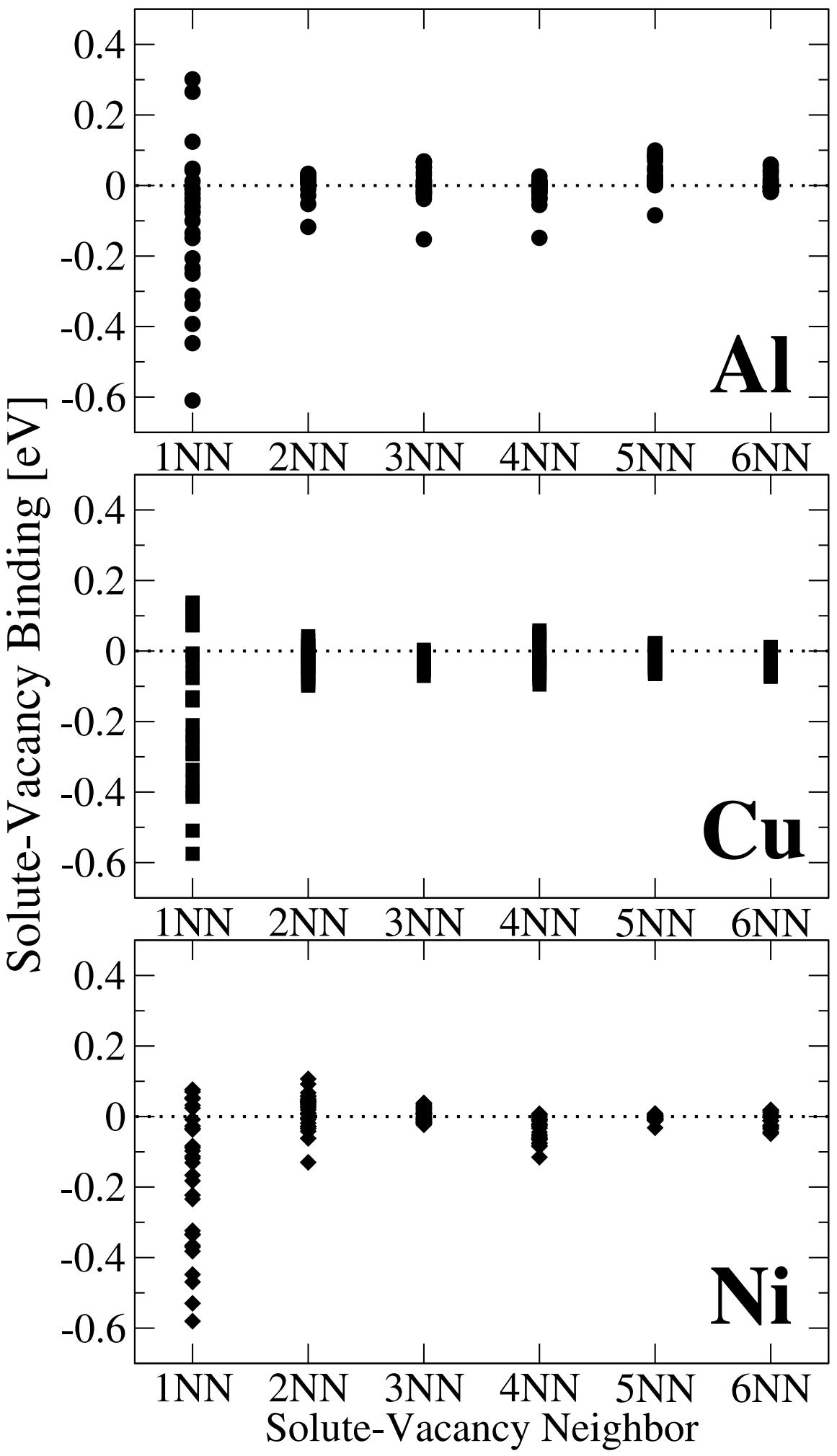

Figure 4: Solute-vacancy binding within Al, Cu, and Ni with respect to neighboring distance. A negative binding indicates an attractive solute-vacancy interaction, and a positive binding indicates a repulsive interaction. Each point represents one solute.

Figure 4 shows solute-vacancy binding energy at up to sixth nearest-neighbor (6NN) separations within Al, Cu, and Ni hosts, where these are the energies to bind the solute and vacancy from effectively infinite separation. We see a large 1NN interaction in all three hosts, followed by mostly less than ±100 meV bindings for all other separations. We calculate the dissociation/association hop as between the 1NN and the 4NN. Therefore, we use the 4NN solute-vacancy binding energy as a measure of the term we have ignored. While it is not clear how to include these long-range binding effects rigorously in the full five-frequency model, we can qualitatively estimate their impact by correcting the energetics of the $\omega_3$ and $\omega_4$ hops so that they are consistent with the energy of complete dissociation. There are many ways to modify the dissociation/association hop barriers to ultimately obtain the correct long distance solute-vacancy binding. We choose to use the kinetically resolved activation (KRA) barrier approximation,[31] which divides the necessary 4NN correction energy in two and applies half to each of the $\omega_3$ and $\omega_4$ barriers. The new $\omega_3$ and $\omega_4$ hops are now reintroduced into the five-frequency model and all solute diffusivities are calculated again. Surprisingly we find that applying this solute-vacancy binding correction gives almost no change, and actually slightly worsens our comparison to experiment through the metric of ($E_{host}^{RMS}$, $E_{host}^{w-RMS}$, $D_{host}^{ratio}$). This shows that the effects of further neighbor solute-vacancy interactions do not have a significant effect on solute diffusivity compared to other sources of error in the systems we have tested, and we therefore assume it is of negligible importance for all the calculations in the present database. We note that some studies on BCC alloys have shown a potentially significant influence of these binding energies on some diffusion phenomena[32].

In calculating the attempt frequency prefactor for each jump in our diffusion model, we only considered the phonon modes of the migrating atom, as this produces a significant timesaving compared to including more atoms. While these modes capture a significant amount of information about changes in the attempt frequency, it assumes that the surrounding atomic phonon modes are not affected by the presence of the solute or vacancy. To assess the impact of the excluded modes, we calculate and plot in Figure 5 the attempt frequencies for Ag, Au, and Cr diffusing in Al when using only the migrating atom, as well as when also considering the nearest 4 atomic neighbors. We see that by using additional phonon modes from surrounding atoms, the calculated attempt frequencies are generally reduced by a factor of two for all frequencies. While a factor of two may be a large error for any particular attempt frequency, a uniform scaling of all attempt frequencies ends up largely cancelling in the five-frequency model, leading to only the same scaling factor on the predicted diffusivity, with no change in the predicted diffusion activation barrier. Also, since $\upsilon_0$, the attempt frequency for the host self-hop, appear to scale the same way as other hops, the accuracy of the predicted D values in this work would not be impacted by this shift as the prefactors are scaled by our DFT/experiment host self-diffusivity fitting correction scheme. Therefore, we conclude that while phonon modes from additional neighboring atoms would produce a more accurate attempt frequency prefactor, it would not significantly improve solute diffusion predictions when our solute diffusion correction method is also being used.

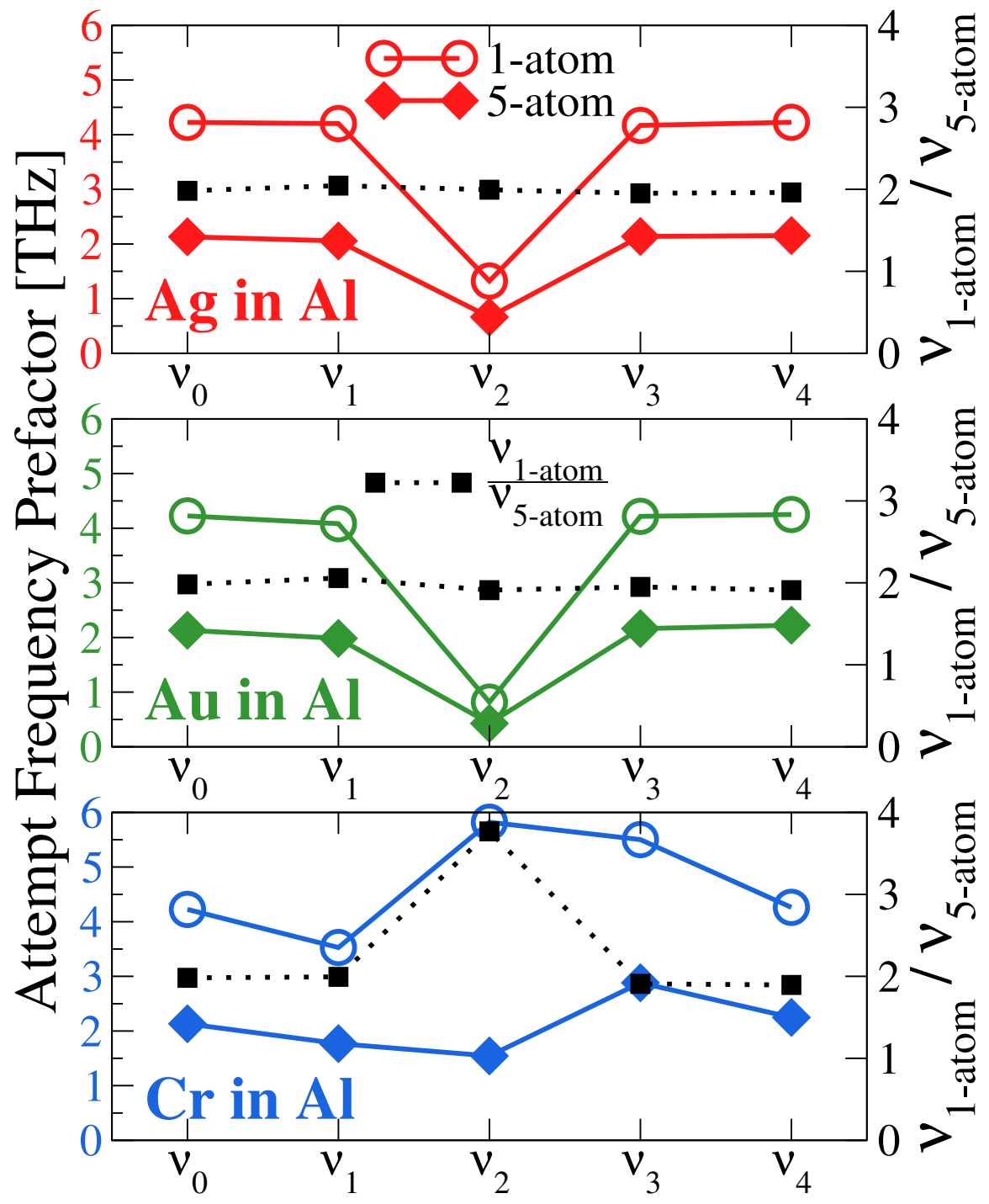

Figure 5: Calculated attempt frequency predictions (left-side y-axis) for Ag, Au, and Cr in Al-host using only the phonon vibrational modes of the migrating atom (colored open symbols) versus using the migrating atom and its four nearest atomic neighbors (colored filled symbols). The dotted lines represent the ratio of the single atom attempt frequency divided by the 5-atom attempt frequency (right-side y-axis). The attempt frequency for each of the five-frequencies is horizontally separated in the plot.

## Usage Notes:

We recommend direct usage of the reported solute diffusion coefficients, $D_0$, and solute diffusion activation energy, $Q$, to generate temperature dependent solute diffusivities. Researchers who would like to instead regenerate the diffusivity data from the reported individual hop barriers and attempt frequencies should remember to apply the host self-diffusivity correction from Table I. In other words, the difference between calculated solute diffusivity and the host self-diffusivity should be the quantity held in high confidence. We recommend caution when using the calculated diffusivity values of magnetic solutes, Cr, Mn, Fe, Co, and Ni in non-magnetic host alloys, as they exhibit much larger errors that our other impurities when compared to experimental measurements.

**Data Citations:**

1. Wu, H., Mayeshiba, T., & Morgan, D. *Figshare* http://dx.doi.org/10.6084/m9.figshare.1546772

**Acknowledgements:**
Funding for this work and the MAST code package were provided by the NSF Software Infrastructure for Sustained Innovation ($SI^2$) award No. 1148011. Tam Mayeshiba was funded by the NSF Graduate Fellowship Program under Grant No. DGE-0718123 and the UW-Madison Graduate Engineering Research Scholars Program. Computational resources for this work came from Extreme Science and Engineering Discovery Environment (XSEDE), the UW-Madison Center For High Throughput Computing (CHTC) and Advanced Computing Initiative (ACI) in the Department of Computer Sciences, and the Lipscomb High Performance Computing Cluster (DLX) at the University of Kentucky Information Technology department.

**Author Contributions:**
H.W. performed most of the diffusion calculations, developed the high-throughput workflows, and worked on data analysis and verification. T.M. performed some of the diffusion calculations and helped develop the MAST tools to automate the workflow. D.M. supervised and planned the work. All authors contributed in writing the manuscript.